\documentclass[prb,aps,twocolumn,showpacs,superscriptaddress]{revtex4-1}

\usepackage{amsmath,amsfonts}
\usepackage{graphicx}
\usepackage{tabularx}

\usepackage{dcolumn}
\usepackage{longtable}
\usepackage{color}
\usepackage{setspace}

\makeatletter
\AtBeginDocument{\@ifpackageloaded{natbib}{\ifNAT@numbers\if@filesw\immediate\write\@auxout{\string\global\string\NAT@numberstrue}\fi\fi}{}}
\makeatother
\begin{document}

\title{Surface States of  Perovskite Iridates AIrO$_3$;\\
Signatures of Topological Crystalline Metal with Nontrivial $\mathbb{Z}_2$ Index}

\author{Heung-Sik Kim}
\affiliation{Department of Physics, University of Toronto, 
Ontario M5S 1A7 Canada}

\author{Yige Chen}
\affiliation{Department of Physics, University of Toronto, 
Ontario M5S 1A7 Canada}

\author{Hae-Young Kee}
\email{hykee@physics.utoronto.ca}
\affiliation{Department of Physics, University of Toronto, 
Ontario M5S 1A7 Canada}
\affiliation{Canadian Institute for Advanced Research, CIFAR Program in Quantum Materials, 
Toronto, Ontario M5G 1Z8, Canada}

\begin{abstract}
There have been increasing efforts in realizing topological metallic phases with nontrivial surface states.
It was suggested that orthorhombic perovskite iridates
are classified as a topological crystalline metal (TCM) with flat surface states protected by lattice symmetries.
Here we perform first-principles electronic structure calculations 
for epitaxially stabilized orthorhombic perovskite iridates.
Remarkably, two different types of topological surface states are found depending on surface directions.
On side surfaces, flat surface states protected by lattice symmetries emerge manifesting the topological crystalline character.
On top surface, on the other hand, an unexpected Dirac cone appears indicating surface states protected by a time-reversal symmetry,
which is confirmed by the presence of a nontrivial topological $\mathbb{Z}_2$ index.
These results suggest that the orthorhombic iridates are unique systems 
exhibiting both lattice- and global-symmetry-protected topological phases and surface states.
Transitions to weak and strong topological insulators and implications of surface states in light of 
angle resolved photoemission spectroscopy are also discussed.
\end{abstract}

\pacs{73.20.At, 71.20.Be, 71.70.Ej}

\maketitle

\section{Introduction}
Recently, a wide range of novel phases including topological Mott insulators, axion insulators, and Kitaev spin liquids
was investigated in correlated electronic systems with strong spin-orbit coupling (SOC)\cite{Review:jeff}. 
Among various candidate materials, iridium oxides have provided an excellent playground to study such combined effects
\cite{Moon,Kim:PRL,Kim:Science,JHKim,Arita,JWKim:327,JHKim2,JMCarter3,Pesin:TMI,Wan:WS,Go:WS,William,JMCarter,AZeb,JMCarter2,Jackeli,JRau}.
In particular, it was suggested that orthorhombic perovskite iridates AIrO$_3$ where A is alkaline earth metal 
fall into a new class of metal dubbed topological crystalline metal (TCM)\cite{CLK}.
TCM is distinguished by flat surface states protected by the crystal symmetry.

It is interesting to note that the bulk electronic band structure of orthorhombic perovskite SrIrO$_3$ was investigated earlier, and
a ring-shape nodal Fermi surface (FS) due to SOC was identified\cite{JMCarter,AZeb}. 
This ring FS is protected by the crystal symmetry, and 
has linear dispersion in both radial and  longitudinal directions perpendicular to azimuthal direction along the ring.
Each point on the ring is composed of two dimensional (2D) Dirac cone. 
It was further shown that breaking 
the mirror symmetry along c-axis leads to a three dimensional (3D) Dirac semimetal or a strong topological insulator 
depending on the strength of broken symmetry\cite{JMCarter}. 

These previous studies imply that surface states may differ depending on surface direction of AIrO$_3$,
since top (or bottom) surface naturally breaks the mirror symmetry. 
Furthermore, the TCM proposal employs an effective single-band model based on the $j_{\rm eff}$=1/2 picture,
which is an assumption that is currently in question in these iridate compounds\cite{Nie_ARPES,Liu_ARPES}.
Further discussion on iridates in general is provided in Sec. \ref{sec:bulkbands}.

To understand topological nature of these iridates and to check the validity of single band model,
we perform first-principles calculations for AIrO$_3$ epitaxially stabilized on different substrates.
In the bulk, we find that the $j_{\rm eff}=1/2$ band is better separated from the 
$j_{\rm eff}=3/2$ states when $U_{\rm eff}$ increases as shown in Fig. 1(b) and (c). 
This suggests that the effective strength of SOC is enhanced by electron interaction, 
supporting the $j_{\rm eff}=1/2$ picture for iridates. On the surfaces,
we observe flat surface states (flat in one-direction, and dispersing linear along the other direction) 
on (110) and ($\bar{1}$10) side surfaces. These are protected by the mirror and glide symmetries, signalling the TCM character.  
On the other hand, on the top surface of (001) direction where the mirror symmetry is broken,
surprisingly a Dirac surface state appears indicating a non-trivial nodal metallic state. 
Indeed, we find that this nodal metal is characterized by a topological $\mathbb{Z}_2$ index 
defined on a 2D side-plane in 3D Brillouin zone, and thus
reinforces the Dirac surface state protected by time reversal symmetry (TRS).

\section{Computational details}
\label{sec:detail}
The results in this study are obtained by using two different density functional theory codes; 
OPENMX and Vienna {\it ab-initio} Simulation Package\cite{OpenMX,VASP1,VASP2}.
For the structural optimization, we used the projector-augmented wave potentials
and the PBEsol\cite{PBEsol} generalized gradient approximation as implemented in 
VASP. 12$\times$12$\times$8 Monkhorst-Pack
grid and 500 eV of plane-wave energy cutoff were used for the momentum space sampling
and the basis set, respectively. The force criterion was 10$^{-3}$eV/\AA. 
For the electronic structure calculations with SOC and on-site Coulomb interaction,
OPENMX code, which is based on the linear-combinaion-of-pseudo-atomic-orbital basis formalism, is used.
A non-collinear DFT scheme and a fully relativistic $j$-dependent pseudopotential are used to treat SOC, and
the parametrization of Perdew and Zunger for the local density approximation was chosen for the exchange-correlation functional\cite{CA,PZ}. 
400 Ry of energy cutoff was used for the real-space integration. 
On-site Coulomb interaction is treated via a simplified DFT+$U$ formalism implemented in OPENMX 
code, and up to 2.0 eV of $U_{\rm eff}\equiv U-J$ parameter were used for Ir $d$ orbital 
in our DFT+SOC+$U$ calculations\cite{Dudarev,OpenMX:LDA+U}. 
Using Maximally-localized Wannier orbital formalism implemented in OpenMX code,
we obtain the hopping parameters for slab Hamiltonian and compute the surface 
band structures\cite{MLWF,MLWF:DISE,OpenMX:Wannier}.

\section{SOC in iridiates and effects of electronic interaction on bulk electronic band structures}
\label{sec:bulkbands}
In transition metal oxides with strong SOC, $t_{\rm 2g}$ states separate from $e_{\rm g}$ states due to
the crystal field split into $j_{\rm eff}$=1/2 and 3/2 states.
As a result, the multiorbital $t_{\rm 2g}$ bands can be mapped to
a single $j_{\rm eff}=1/2$ band for $d^{5}$ valence configuration.
Since the bandwidth of $j_{\rm eff}=1/2$ states is significantly narrower 
than the entire $t_{\rm 2g}$ bandwidth, the effect of electron interaction 
is amplified in the $j_{\rm eff}=1/2$ band. This was found to be the origin 
for the insulating and magnetic behavior observed in quasi-two-dimensional layered iridates
Sr$_2$IrO$_4$ and Sr$_3$Ir$_2$O$_7$\cite{Kim:PRL,Moon,JHKim2}, which was naively
expected to be metallic due to significantly weaker on-site Coulomb interaction 
of $U \sim 2$ eV in the iridium 5$d$ orbital\cite{Arita,cRPA2}.

Contrary to Sr$_2$IrO$_4$ and Sr$_3$Ir$_2$O$_7$, SrIrO$_3$ shows qualitative differences.
SrIrO$_3$ is reported to be a nonmagnetic semimetal\cite{Cao_SIO}. 
It was proposed that the semimetallic behavior is due to the nodal ring-shaped 
FS\cite{JMCarter,AZeb}.
It was later found that, the nodal FS protected by crystal symmetry is
a signature of the TCM phase\cite{CLK}.
Since these studies are based on $j_{\rm eff}=1/2$ single band model,
it is important to determine the validity of $j_{\rm eff}=1/2$ picture in SrIrO$_3$, 
as $j_{\rm eff}=3/2$ states can be mixed with $j_{\rm eff}=1/2$ states 
as well as the effect of electron interaction to the semimetallic. Below we present 
our {\it ab-initio} calculation results for bulk and surface states.


We study two AIrO$_3$ on two different substrates;
orthorhombic perovskite phase of CaIrO$_3$ and SrIrO$_3$ stabilized on GdScO$_3$ (a=3.967\AA)
and SrTiO$_3$ (a=3.905\AA) substrates with $Pbnm$ space group symmetry (No. 62). 
For each system, c-axis lattice constant and internal atomic coordinates are determined 
from the structural optimization. 
Since all of the systems share qualitatively same features,  
hereafter we show the results of CaIrO$_3$ on GdScO$_3$ substrate as a representative example,
which shows the most prominent surface states. 
The crystal structures and bands for the other three compounds are presented in Sec. \ref{sec:bulk}
in the Appendix.

\begin{figure}
  \centering
  \includegraphics[width=0.48\textwidth]{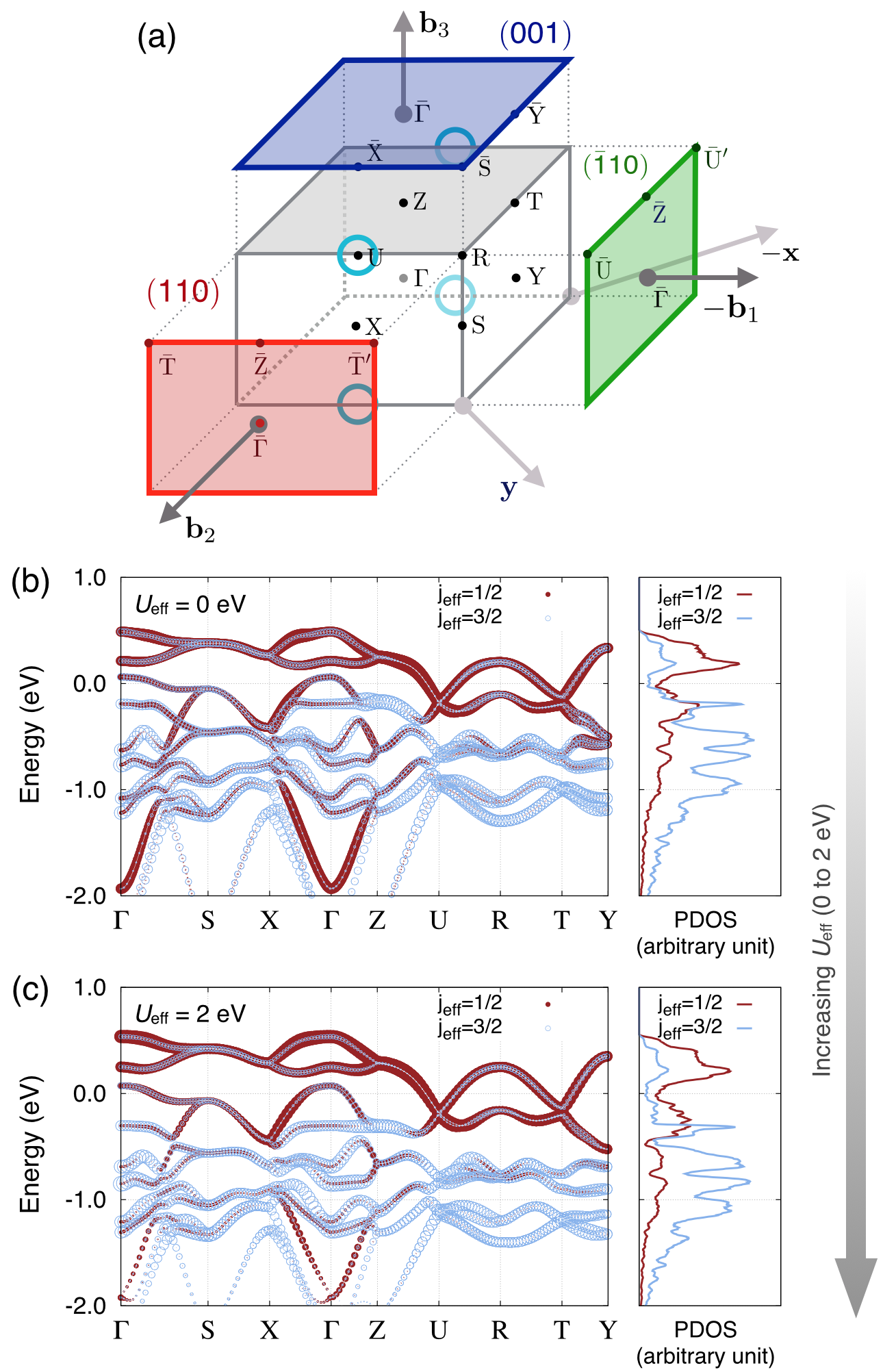}
  \caption{(Color online)
  (a) Special $k$-points in the bulk and on the surfaces. 
  Note that $\mathbf{b}_i \parallel \mathbf{a}_i~(i=1,2,3)$,
  where $\mathbf{a}_i$ are the Bravais lattice vectors. $\mathbf{x}$, $\mathbf{y}$, and
  $\mathbf{z}\equiv \mathbf{x} \times \mathbf{y}$ are the
  local coordinates of IrO$_6$ octahedra in the absence of octahedral rotation.
  Bulk nodal lines are depicted as cyan circles near U points (size is exaggerated). 
  Bulk $j_{\rm eff}$-projected band structure and density of states 
  of CaIrO$_3$ on GdScO$_3$ substrate with (b) $U_{\rm eff}$ = 0 and (c) 2 eV. 
  The weight of $j_{\rm eff}$=1/2 and 3/2 states are denoted as
  the thickness of dark blue lines and size of red circular symbols, respectively.
}
  \label{fig:bulk}
\end{figure}

Fig. \ref{fig:bulk} shows the $j_{\rm eff}$-projected electronic structures
of CaIrO$_3$ on GdScO$_3$ substrate in the presence of SOC,
where
the bulk and surface Brillouin zones (BZ) are depicted in Fig. \ref{fig:bulk}(a), and
Fig. \ref{fig:bulk}(b) and (c) show the band structure without and with on-site Coulomb interaction, respectively.
The effective Coulomb interaction parameter $U_{\rm eff}\equiv U-J$ (where $J$ is Hund's coupling) is increased up to 2 eV,
and the paramagnetic semimetallic phase with the nodal line remains unchanged. 
One can see a cut of the nodal ring close to ${\rm U}$-point where
the location of the ring is shown as cyan circles in Fig. \ref{fig:bulk}(a).
Dark red and blue colors refer to $j_{\rm eff}$=1/2 and 3/2, respectively, and the size of circles
represents their weight.   

In the top four bands $j_{\rm eff}$=1/2 character is dominant, but there are non-negligible amount of
$j_{\rm eff}$=3/2 states mixed into the bands over the BZ, which is also shown in the 
projected density of states (PDOS).  However, it is important to note that the bands near U-point where the nodal ring
exists is mainly composed of $j_{\rm eff}$=1/2, validating the effective single band model for the nodal FS used in Ref. \onlinecite{CLK}.
%
By comparing Fig. \ref{fig:bulk}(b) and (c), the $j_{\rm eff}$=1/2 states are
pushed up and enhanced near the Fermi level, and on the contrary, 
the $j_{\rm eff}$=3/2 states are pushed down in the presence of $U_{\rm eff}$,
even though the nodal ring composition is minimally affected. 
Comparing the PDOS with nonzero $U_{\rm eff}$ to that with artificially increased SOC,
inclusion of $U_{\rm eff}$ = 2 eV corresponds to approximately 30\% enhancement of SOC. 
Such behavior can be understood as the enhancement of orbital polarization in terms of 
spin-orbital entangled basis, as reported in a previous LDA+$U$ study on Sr$_2$RhO$_4$\cite{Liu_SRO}. 
In other words, fully occupied $j_{\rm eff}$=3/2 
states are energetically favored in the presence of both $U_{\rm eff}$
and sizable SOC\cite{Dudarev,OpenMX:LDA+U}. 
Unless the Hund's coupling becomes a dominant factor, such effect should persist
in more general treatment of Coulomb interaction. For cross-checking purpose we performed 
another LDA+$U$ calculations using Lichtenstein's generalized LDA+$U$ formalism 
as implemented in VASP code\cite{Lich}, 
which gives consistent results for $U\leq$ 2.0 eV and $J\leq$ 0.5 eV. 
It should be noted that, such behavior is also observed in recent dynamical mean-field theory 
calculations\cite{ZHV,Arita,sato_t2gSOC}, indicating that the enhancement of SOC by electronic interaction is
a general feature of correlated electronic systems with SOC.

\begin{figure}
  \centering
  \includegraphics[width=0.48\textwidth]{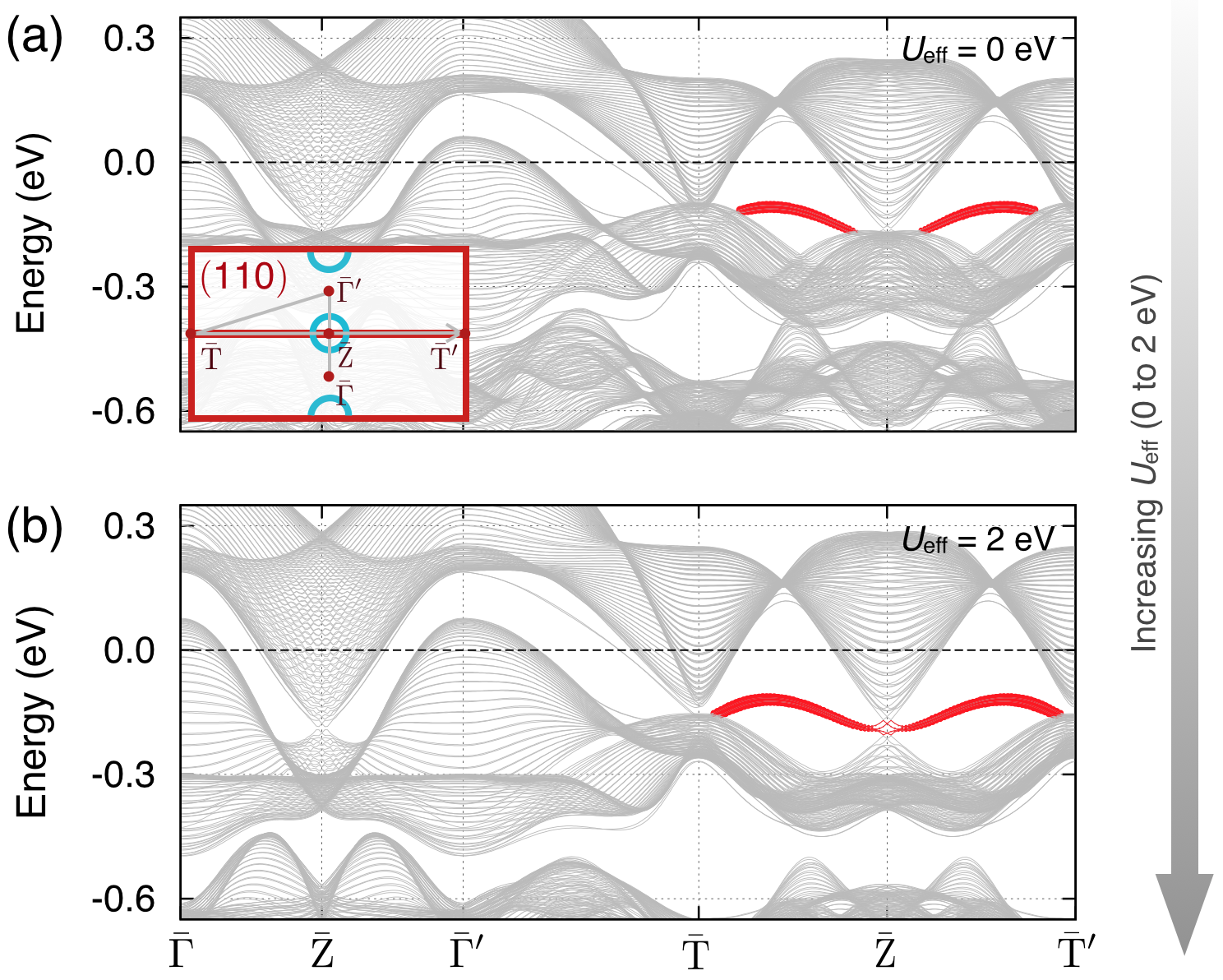}
  \caption{(Color online)
    Band structure of a slab geometry parallel to (110) surface (30 unit cell thickness)
    with (a) $U_{\rm eff}$ = 0 and (b) 2 eV. 
    The thickness of red lines overlaid with the bands represents surface weight.
}
  \label{fig:surface1}
\end{figure}

In the following sections we consider surface states. In this work we employ three surface directions;
($\bar{1}$10), (110), and (001) surfaces in terms of the local coordinates, which are perpendicular to
$\mathbf{a}_1$, $\mathbf{a}_2$, and $\mathbf{a}_3$ as shown in Fig. \ref{fig:bulk}(a), respectively.
Among them, (110) and ($\bar{1}$10) contain the mirror symmetry, while (001) surface does not. 
For each cases, we employed supercells having thickness of 30 unit cells along the surface normal directions.
Results on other sides such as (100) and (010) orientations are discussed 
in Sec. \ref{sec:surface_alt} in the Appendix.

\section{Flat Surface States on Side Surfaces with Mirror Symmetry}
Fig. \ref{fig:surface1}(a) and (b) show the band structure of (110) slab with 
$U_{\rm eff}$ = 0 and 2 eV, respectively. 
To visualize which band originates from the surface, each Bloch state is projected onto the 
atomic orbitals located on the surface in the slab geometry. 
Surface weight, which is the size of surface-projected probability amplitude in each Bloch state,
is shown as thickness of the colored lines overlaid in the figures. 
In both plots, the surface states are clearly present slightly below the Fermi level,
which are nearly dispersionless along the $\bar{\rm T}$-$\bar{\rm Z}$-$\bar{\rm T}'$ where $k_z=\frac{\pi}{c}$,
while  they show linear dispersion along $\Gamma-\bar{\rm Z}-\Gamma'$ ($c$-direction).
The presence of these flat surface states is a manifestation of  bulk TCM phase.
Our result suggests that the single band $j_{\rm eff}=1/2$ tight binding model used in Ref. \cite{CLK} is valid even though
the bands near Fermi level involve $j_{\rm eff}=3/2$ states. This is because the mixing of $j_{\rm eff}=3/2$ states does not alter 
the topology of this nodal metal.

However, ideally the surface states perpendicular to ${\bf b}_2$ for a fixed $k_z=\frac{\pi}{c}$ should be completely flat 
along the ${\hat b}_1$-direction,
because of the presence of bulk chiral  symmetry \cite{CLK}; note that there are four sublattices in perovskite iridates leading to a
sublattice chiral symmetry. Slight dispersion of this surface state is due to the presence of a small 
chiral symmetry breaking term such as a finite hopping integral between the same sublattice in the $a-b$ plane.
The mixing of the $j_{\rm eff}$=3/2 states into the $j_{\rm eff}$=1/2 bands further breaks the chiral symmetry,
but its effect is minor compared to the same sublattice hopping integral effects.
Introducing electronic interaction pushes down the bulk $j_{\rm eff}$=3/2 states, and thus
reveals the surface states more visible.  The linear dispersing surface state
along the $\Gamma-\bar{\rm Z}-\Gamma'$ exists, but its weight is too small as it mixes with the bulk states.

\begin{figure}
  \centering
  \includegraphics[width=0.48\textwidth]{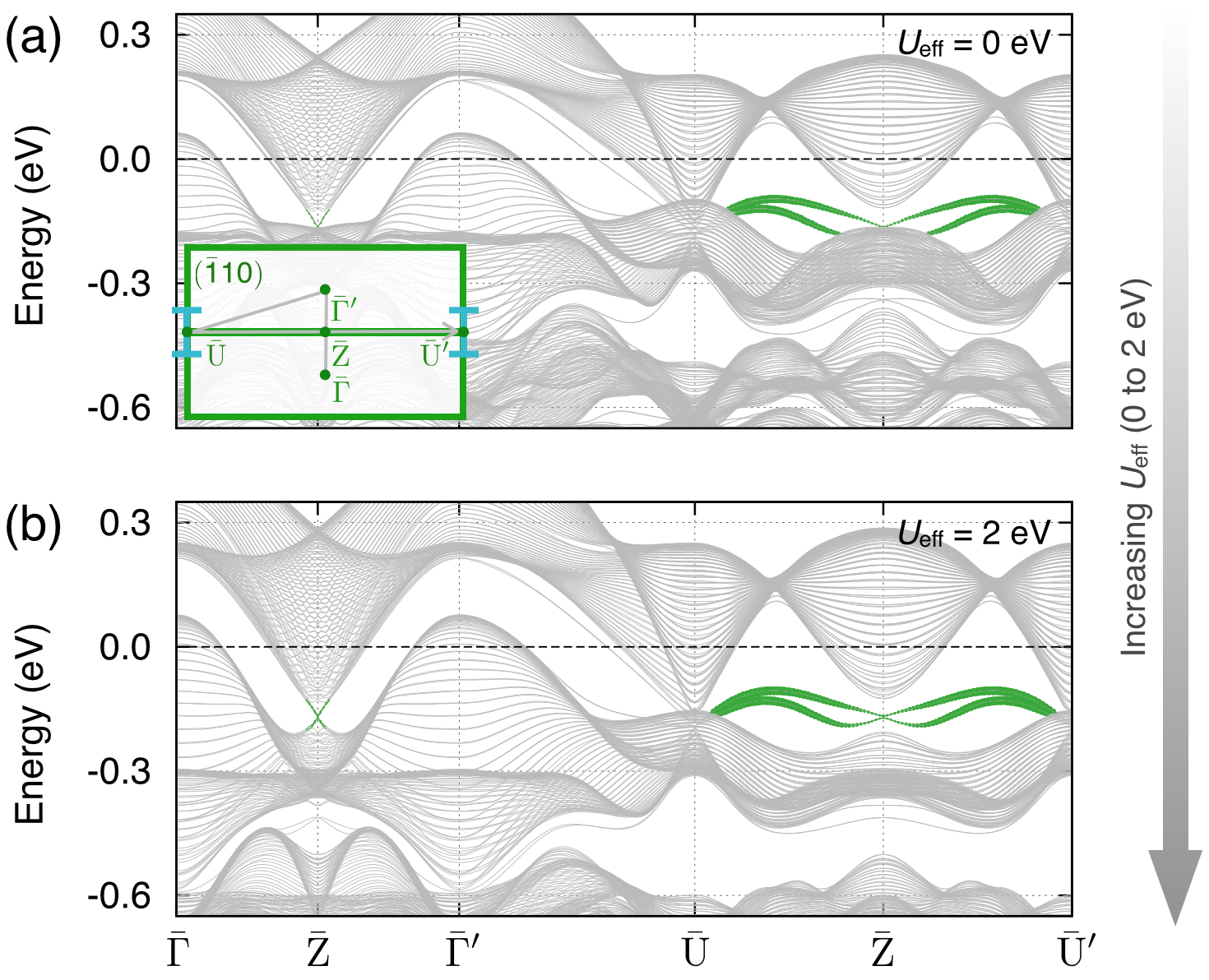}
  \caption{(Color online)
    Band structure of a slab geometry parallel to ($\bar{1}$10) surface (30 unit cell thickness)
    with (b) $U_{\rm eff}$ = 0 and (c) 2 eV.
    The thickness of green lines overlaid with the bands represents surface weight.
}
  \label{fig:surface2}
\end{figure}

Fig. \ref{fig:surface2} shows the band structure of ($\bar{1}$10) slab,
depicted as the green side in Fig. \ref{fig:bulk}(a).
The band structure with $U_{\rm eff}$ = 0 and 2 eV are shown in Fig. \ref{fig:surface2}(a) and (b), respectively.
The separation of the bulk and surface bands owing to electronic interaction is also found.
Unlike the (110) surface, bulk spectrum is gapped at $\bar{\rm Z}$ point
due to the absence of nodal ring projected onto ($\bar{1}$10) plane.
Thus the linear dispersing surface state along $\Gamma-\bar{\rm Z}-\Gamma'$ is observable and
denoted by a light green color. 

Note that, the chiral symmetry is also lifted, as the surface states split into two from $\bar{\rm U}$ to $\bar{\rm Z}$, 
but the degeneracy is recovered at $\bar{\rm Z}$ due to  
the presence of the  $n$-glide symmetry $\Pi_n$ on the $(\bar{1}10)$ surface in addition to TRS.
The one-dimensional (1D) Hamiltonian, parametrized by the surface momentum $k_{b_2}$, is applicable to represent 
the system with open $(\bar{1}1 0)$ surface at $k_z=\frac{\pi}{c}$. The $n$-glide operation only 
reverses the sign of $k_{b_2}$ when acting on the 1D Hamiltonian, which suggested that $\Pi_n$ can be 
treated effectively as a spatial inversion operator. 
Accordingly, the degeneracy at ${\bar{\rm Z}}$,
which is a $n$-glide-symmetry-invariant momentum, is protected by the n-glide symmetry and TRS.

\begin{figure}
  \centering
  \includegraphics[width=0.48\textwidth]{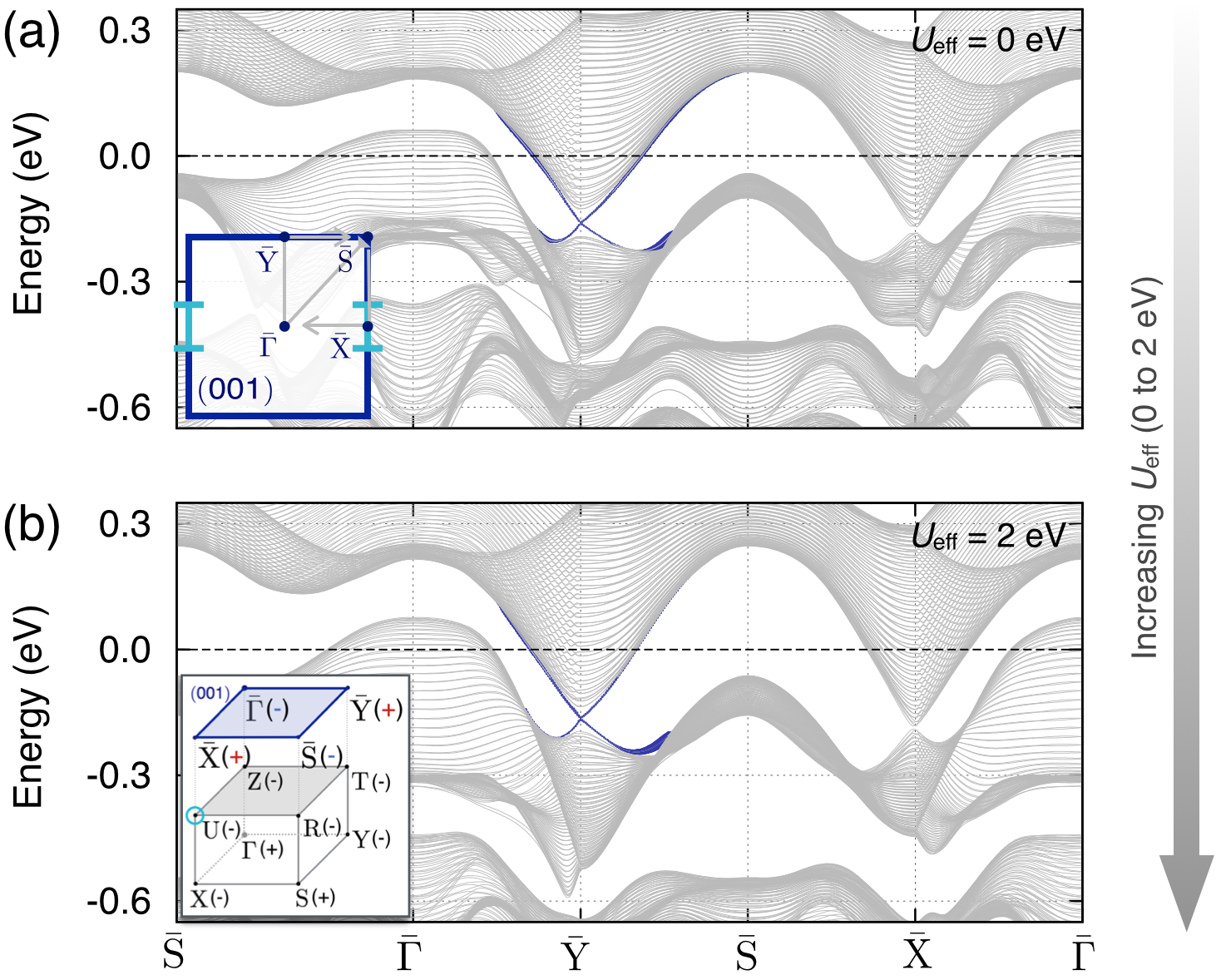}
  \caption{(Color online)
    Band structure of a slab geometry parallel to (001) surface (30 unit cell thickness)
    with (b) $U_{\rm eff}$ = 0 and (c) 2 eV. 
     The thickness of dark blue lines overlaid with the bands represents surface weight.
    Inset in (b) shows the product of the parity eigenvalues of valence $j_{\rm eff}$=1/2 bands
    on each TRIM points in the bulk and the (001) surface Brillouin zones.
}
  \label{fig:surface3}
\end{figure}

\section{Surface Dirac Cone on Top Surface}
Now let us examine the (001) surface. Due to the loss of mirror symmetry, the theory of TCM does not apply here, and one does not
expect any protected surface states on the (001) plane\cite{CLK}. 
However,  as shown in Fig. \ref{fig:surface3}, a Dirac cone at
$\bar{\rm Y}$ point emerges, which carries rather weak but still finite surface weight. 
This surface state is an indication of a non-trivial bulk topological $\mathbb{Z}_2$ invariant. 
While the 3D iridate is a semi-metal with the nodal ring FS,  the 2D bands in $k_{b_1}=\frac{\pi}{a}$ plane have a gap across
the Fermi level, and thus there exists a well-defined 2D topological $\mathbb{Z}_2$ index, $\nu_{2D}$.
This $\nu_{2D}$ can be obtained by multiplying the parity eigenvalues at each TRIM point in $k_{b_1}=\frac{\pi}{a}$ plane shown in the inset of Fig. \ref{fig:surface3}(c).  We find that $\nu_{2D}=1$ which implies the presence of the 2D surface Dirac cone at $\bar{\rm Y}$ point. \cite{FuKane,YangNagaosa}. 
On the other hand, due to the existence of the gapless Dirac bulk points in $k_{b_1}=0$ plane, we do not expect protected surface states at $\bar{\rm X}$ point,  which is indeed shown in Fig \ref{fig:surface3}.  

This nontrivial topology of nodal metal leads to other topological insulators via phase transition by lowering the  crystal symmetry.
For example, a weak topological insulator (WTI) can be realized by adding a small 
perturbation that does not invert the bands but breaks the mirror and inplane sublattice symmetries. 
This WTI has the weak $\mathbb{Z}_2$ indices, ($\nu_1$$\nu_2$$\nu_3$) = (110)
and exhibits two surface Dirac cones at both $\bar{\rm X}$ and $\bar{\rm Y}$ points (see Sec. \ref{sec:wti} in the Appendix). 
When the mirror symmetry breaking is large enough to invert the parity at R-point, the system becomes a strong topological insulator
as reported in Ref. \cite{JMCarter}.
%
%

Remarkably, this iridate can hold both the crystal symmetry protected flat surface states (thus named topological crystalline metal), and TRS protected Dirac surface state indicated by a 2D topological $\mathbb{Z}_2$ index.

\section{Discussion and Summary}
While various topological insulators including $\mathbb{Z}_2$ and crystalline insulators are intensively studied,
less efforts have been made to explore topological metallic states. Since a metal with large FS smears surface states by mixing
them with bulk gapless excitations, it is not meaningful to investigate topological surface states. However, a non-trivial semimetal
with small FS pockets, lines, or points can exhibit topological surface states separated from its bulk spectrum.  Here we show
that perovskite iridates possess such phenomena. 
Moreover, these iridates are unusual as different types of topological surface states can be realized depending on the surface direction.
While utilizing epitaxial growth of thin film or superlattices of 
CaIrO$_3$ or SrIrO$_3$ is challenging, there have been several reports on successful growth of perovskite iridates
on various substrates\cite{Jang,Seo1,Seo2,Seo3,Matsuno}. 
Angle resolved photoemission spectroscopy measurement will reveal a novel metallic state, as 
the dispersion of these surface states differ depending on momentum direction in the BZ.


Indeed, a couple of ARPES results on SrIrO$_3$ thin film has been reported recently\cite{Nie_ARPES,Liu_ARPES}.
Both studies show simlar Fermi surface topology at $k_z=\pi / c$ plane (ZURT plane in the bulk) with large
hole-like and small electron-like pockets around R and U (and T) points, respectively, which are also comparable to
our results for SrIrO$_3$ on SrTiO$_3$ substrate shown in Fig. S1(d) in Supplementary Material.
Especially, both experimental results shows the linearly dispersing bands at the U and T points,
which are the signal of the nodal ring in this system as shown in the {\it ab-initio} bands.
Since the {\it ab-initio} results reasonably matches with the ARPES measurements, one can expect to 
observe the surface Dirac cone on the (001) surface in this material as shown in our calculations. Accordingly, further 
studies distinguishing the bulk and surface spectrum are imperative.


In summary,  we propose that the orthorhombic perovskite iridate is a topological semimetal characterized by both 
time reversal and crystal symmetry.
Thus it hosts two distinct types of topologically nontrivial surface states;
the flat surface states protected by the crystal symmetry,  and the Dirac surface state protected by TRS
associated with the 2D topological $\mathbb{Z}_2$ index. The electronic interaction
cooperates with SOC, making such surface states more evident. Considering the recent successes in growing
high-quality samples, these perovskite iridates could be an excellent platform in 
searching for uncovered topological metallic phases.

\begin{acknowledgments}
This work was supported by the NSERC of Canada and the center 
for Quantum Materials at the University of Toronto.
Computations were mainly performed on the GPC supercomputer at the SciNet HPC Consortium.
SciNet is funded by: the Canada Foundation for Innovation under the auspices 
of Compute Canada; the Government of Ontario; Ontario Research Fund - Research Excellence; 
and the University of Toronto.
HSK thanks to IBS Center for Correlated Electron System
in Seoul National University for additional computational resources.
\end{acknowledgments}

\appendix

\begin{table*}
\centering
\begin{tabular*}{\textwidth}{llrrrrrrrrrrrr} \hline\hline \noalign{\medskip}
&~~~~~& \multicolumn{3}{r}{~~~~~CaIrO$_3$ on GdScO$_3$} & \multicolumn{3}{r}{~~~~~CaIrO$_3$ on SrTiO$_3$} & \multicolumn{3}{r}{~~~~~SrIrO$_3$ on GdScO$_3$} & \multicolumn{3}{r}{~~~~~SrIrO$_3$ on SrTiO$_3$} \\
\noalign{\medskip}\hline\noalign{\medskip}
\multicolumn{2}{l}{$a$ (\AA)} & \multicolumn{3}{r}{5.610    } & \multicolumn{3}{r}{5.523    } & \multicolumn{3}{r}{5.610    } & \multicolumn{3}{r}{5.523}     \\ \noalign{\smallskip}
\multicolumn{2}{l}{$c$ (\AA)} & \multicolumn{3}{r}{7.483    } & \multicolumn{3}{r}{7.568    } & \multicolumn{3}{r}{7.817    } & \multicolumn{3}{r}{7.974}     \\ 
\noalign{\medskip} \hline \noalign{\medskip}
         &&$x$&$y$&$z$&$x$&$y$&$z$&$x$&$y$&$z$&$x$&$y$&$z$ \\ \noalign{\medskip}
Ca/Sr &$4c$&~~~~~ -0.013 & -0.055 & 0.250       &~~~~~ -0.013 &  0.444 &  0.250      &~~~~~ -0.007 & -0.039 &  0.250      &~~~~~ -0.007 &  0.461 &  0.250 \\ \noalign{\smallskip}
Ir    &$4a$&~~~~~  0.500 &  0.000 & 0.000       &~~~~~  0.000 &  0.000 &  0.000      &~~~~~  0.500 &  0.000 &  0.000      &~~~~~  0.000 &  0.000 &  0.000 \\ \noalign{\smallskip}
O(1)  &$4c$&~~~~~  0.111 &  0.545 & 0.250       &~~~~~  0.111 &  0.043 &  0.250      &~~~~~  0.081 &  0.524 &  0.250      &~~~~~  0.076 &  0.017 &  0.250 \\ \noalign{\smallskip}
O(2)  &$8d$&~~~~~  0.801 &  0.199 & 0.054       &~~~~~  0.803 &  0.697 &  0.054      &~~~~~  0.790 &  0.210 &  0.042      &~~~~~  0.795 &  0.705 &  0.040 \\ \noalign{\medskip}\hline\hline
\end{tabular*}
\label{tab:coord}
\caption{Lattice constants and internal coordinates of AIrO$_3$ (A=Ca,Sr) on GdScO$_3$ and SrTiO$_3$ substrates.}
\end{table*}

\begin{figure*}
  \centering
  \includegraphics[width=0.95\textwidth]{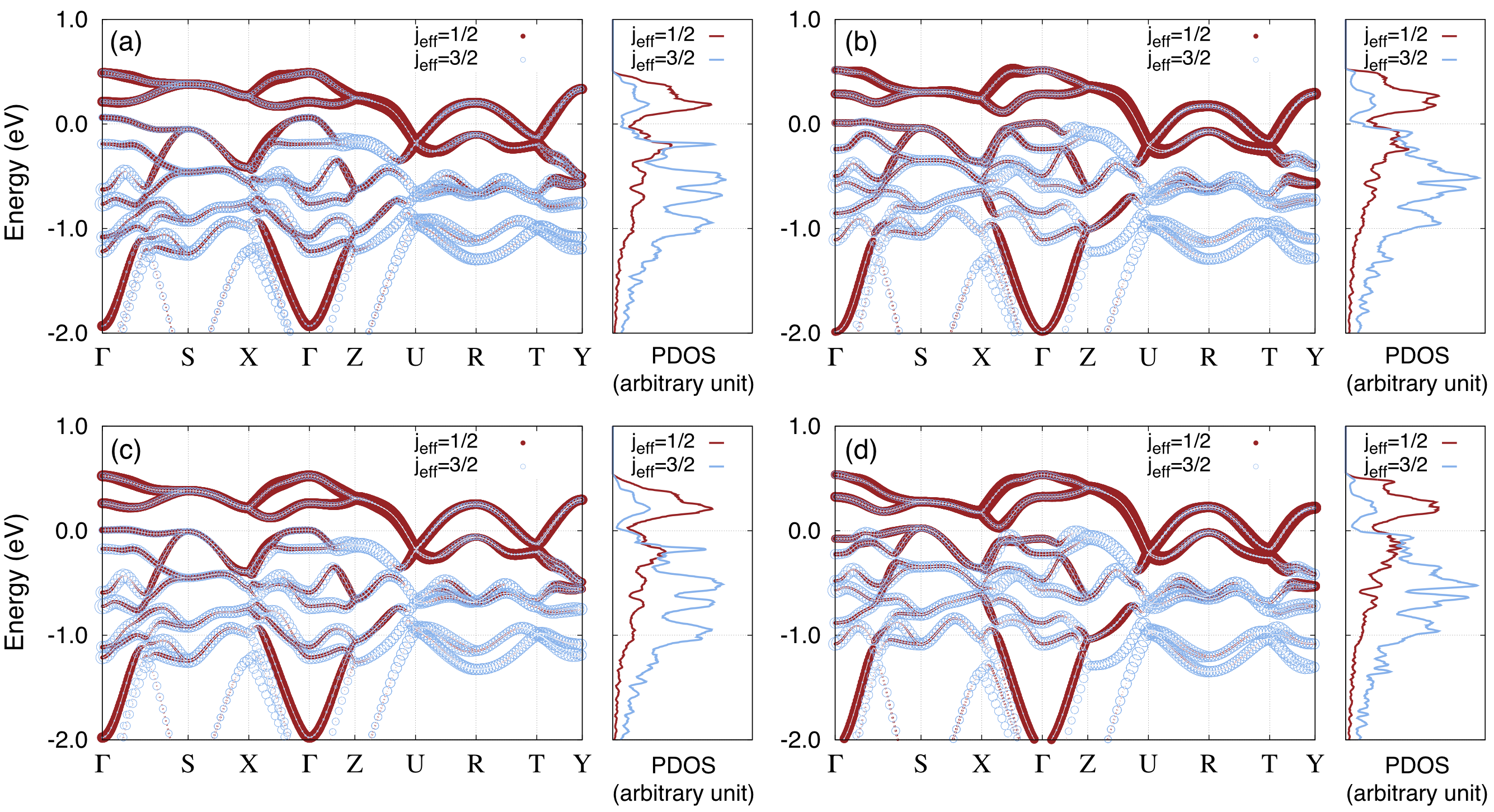}
  \caption{(Color online)
    Band structure of (a) CaIrO$_3$ on GdScO$_3$, (b) SrIrO$_3$ on GdScO$_3$, (c) CaIrO$_3$ on SrTiO$_3$, 
    and (d) SrIrO$_3$ on SrTiO$_3$ in the presence of SOC. 
}
  \label{fig:bulk2}
\end{figure*}

\section{Crystal and electronic structures for all compounds}
\label{sec:bulk}

In this part we show the crystal and electronic structures for CaIrO$_3$ and SrIrO$_3$ on 
GdScO$_3$ and SrTiO$_3$ substrates. All of the systems have $Pbnm$ space group (No. 62) symmetry,
where the lattice parameters and internal coordinates from our calculations are shown in Table \ref{tab:coord}.
Fig. \ref{fig:bulk2} shows the band structure for each compound in the presence of SOC,
where they share similar shape of dispersion as well as the line nodes near U point. 
Comparing (a) to (b) and (c) to (d), SrIrO$_3$ cases have larger bandwidth than CaIrO$_3$ cases
due to the smaller rotation and tilting angles of IrO$_6$ octahedra. 
The change of substrate also yields a slight difference in the bandwidth.

\section{Surface states for different side surfaces}
\label{sec:surface_alt}

\begin{figure}
  \centering
  \includegraphics[width=0.45\textwidth]{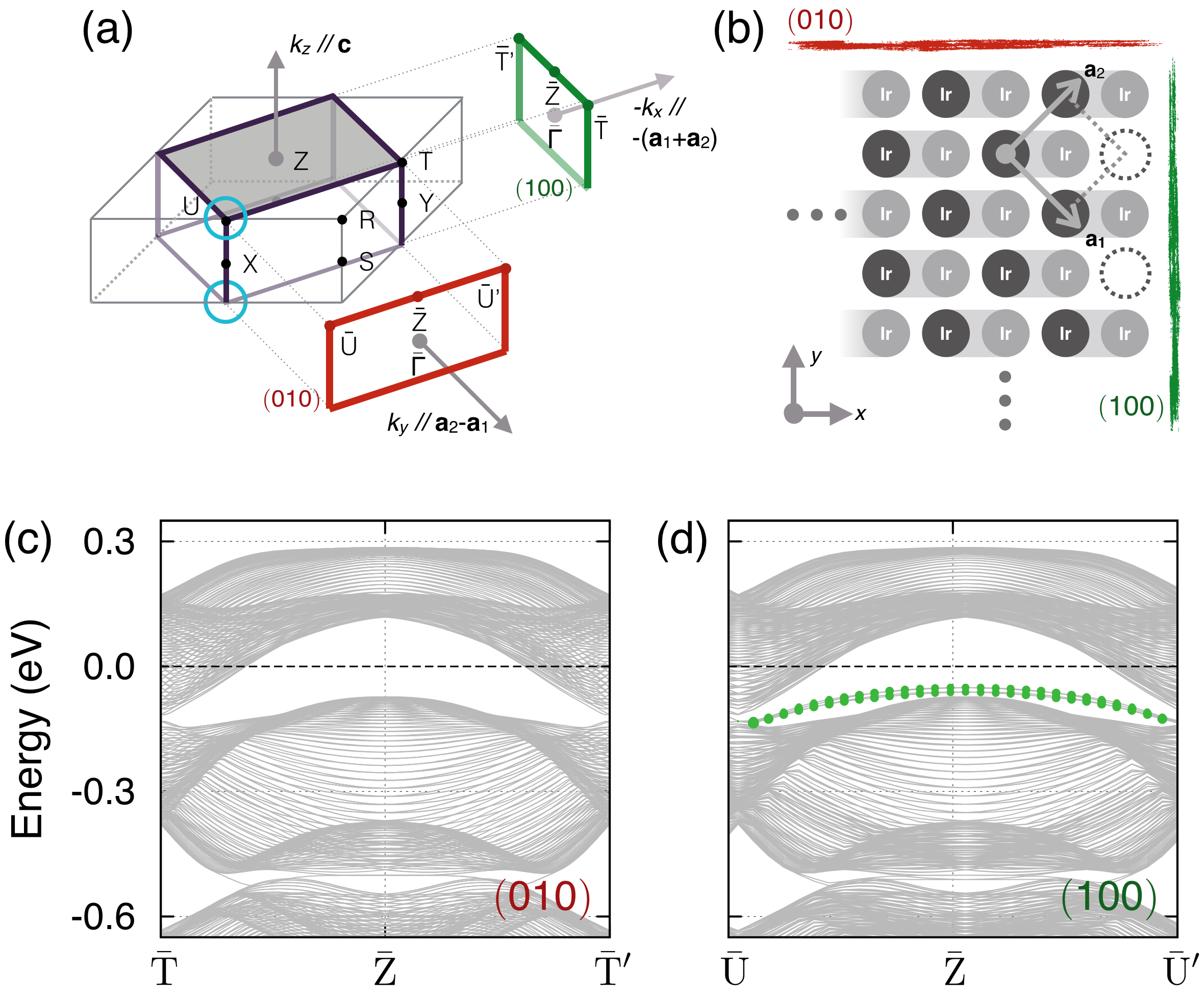}
  \caption{(Color online)
    (a) Surface Brillouin zones for (100) and (010) surfaces and the special $k$ points. 
    (b) Surface configurations for (100) and (010) surfaces, where the unit cell used in the 
    construction of both surfaces is depicted as dashed square.
    The band structures for (c) (100) and (d) (010) slabs.
}
  \label{fig:surface_alt}
\end{figure}

In this section, the surface states on the pseudocubic (100) and (010) surfaces are considered.
The orientation of each surface is shown in Fig. \ref{fig:surface_alt}(a), 
and the configurations for (100) and (010) surfaces with given choice of unit cell is shown in Fig. \ref{fig:surface_alt}(b).
Since the unit cell should contain both sublattices, open boundaries on (100) and (010) surfaces are different
as shown in Fig. \ref{fig:surface_alt}(b). 
The results of slab calculations are shown in Fig. \ref{fig:surface_alt}(c) and (d). 
In the case of (100) surface, there is a flat surface state (denoted as green dots) across the $\bar{\rm U}-\bar{\rm Z}-\bar{\rm U}'$, corroborating 
the results obtained based the winding number analysis of the effective $j_{\rm eff}=1/2$ model in Ref. \onlinecite{CLK}.

\section{Weak topological insulator phase in the absence of mirror and sublattice symmetries}
\label{sec:wti}

\begin{figure}
  \centering
  \includegraphics[width=0.45\textwidth]{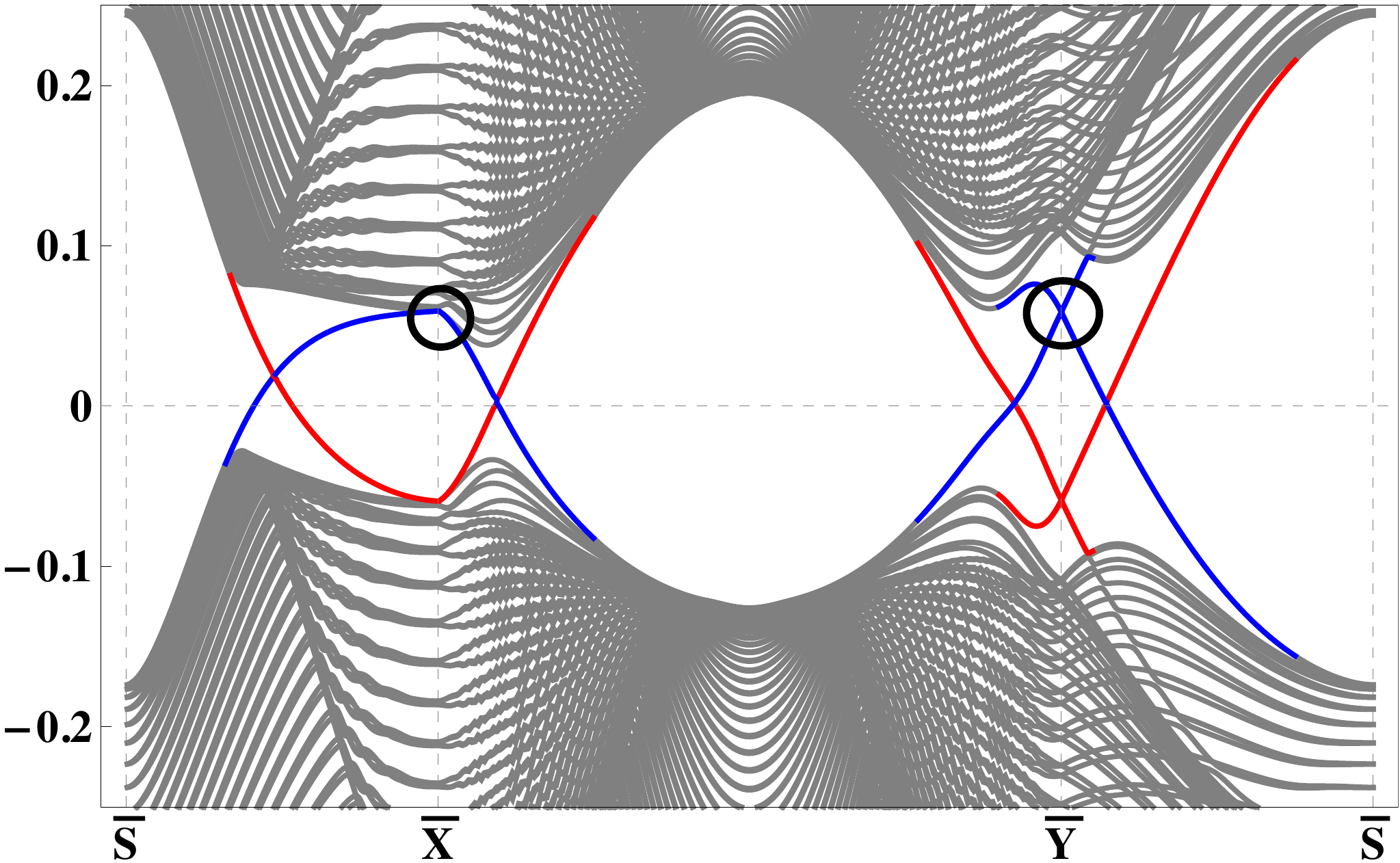}
  \caption{(Color online)
    Band structure of (001) slab based on the tight-binding model in Ref. \onlinecite{CLK}
    with additional terms breaking the mirror and sublattice symmetries.
    Surface states localized on top and bottom surfaces are depicted as blue and red lines,
    respectively. The Dirac cones at top surface are marked by the black circles.
}
  \label{fig:surfacetop}
\end{figure}


As we discussed in the main text, there exists a 2D topological $\mathbb{Z}_2$ index, $\nu_{2D}$ in $k_{{\bf b}_1}$=$\pi/a$ plane.
Due to this nontrivial topology of the nodal metal, one expects that a $\mathbb{Z}_2$ weak or strong topological insulator can be achieved
by breaking crystal symmetries. As shown in Ref. \onlinecite{JMCarter}, a strong mirror symmetry breaking term leads
to a strong topological insulator by inverting the bands at R-point. 
Here we show that a weak topological insulator can be also achieved when a small perturbation that 
breaks both mirror and in-plane sublattice symmetries is introduced. Using the tight binding model 
in Ref. \onlinecite{CLK}, we found that the nodal ring is fully gapped when the mirror and in-plane 
sublattice symmetries are broken. The perturbation is small enough that the parity eigenvalues remain the same.
Thus the $\mathbb{Z}_2$ index is found to be $(\nu_0;\nu_1\nu_2\nu_3)=(0;110)$ from the parity eigenvalues shown
in the inset of Fig. 4(c), indicating a weak topological insulator. 
Fig. \ref{fig:surfacetop} shows the band structures of (001) slab obtained from the tight-binding model
with small mirror and sublattice symmetry breaking terms. 
A finite bulk gap at $\bar{\rm X}$ point is seen and two pairs of gapless surface modes emerge.  
Blue and red lines show the surface states on top and bottom of the slab, respectively.
One can see two Dirac cones at $\bar{\rm X}$ and $\bar{\rm Y}$ points on the top surface 
(marked with black circles in Fig.~\ref{fig:surfacetop}) 
as expected from the weak $\mathbb{Z}_2$ index\cite{FuKane}. Note that, the Dirac points at $\bar{\rm X}$ point are located
at the top and bottom edges of the bulk spectrum, as shown in Fig. \ref{fig:surfacetop},
since the size of bulk gap at $\bar{\rm X}$ point is same with the the energy shift of the Dirac cones 
originating from the symmetry-breaking perturbations.

\bibliography{AIrO3_HSK}{}
\bibliographystyle{apsrev4-1}

\end{document}